\documentstyle[psfig]{aipproc}

\begin{document}
\title{Dark matter from Affleck-Dine baryogenesis
}

\author{Alexander Kusenko}
\address{Department of Physics and Astronomy, University of California, Los
Angeles, CA 90095-1547}



\maketitle

\begin{abstract}
Fragmentation of the Affleck-Dine condensate into Q-balls could fill the
Universe with dark matter either in the form of stable baryonic 
balls, or LSP produced from the decay of unstable Q-balls.  The dark matter
and the ordinary matter in the Universe may share the same origin.
\end{abstract}

The Affleck-Dine (AD) scenario for baryogenesis~\cite{ad} has several
appealing features.  First, the requisite scalar degrees of freedom with a
non-zero baryon number are necessarily present in theories with low-energy
supersymmetry.  Second, the scalar potential of the MSSM and other
supersymmetric generalizations of the Standard Model has numerous flat
directions.  At the end of inflation, the scalar fields develop large
expectation values along these flat directions setting the stage for the AD
scenario.  The most economical supersymmetric extension of the Standard
Model, MSSM, with some appropriate supersymmetry breaking~\cite{drt1}, is
already sufficient for the AD baryogenesis~\cite{drt}.  Finally, the
observed value of the baryon asymmetry, $\eta_{_B}\sim 10^{-10}$, can
easily be accommodated in this scenario.

Supersymmetry also predicts the existence of non-topological solitons,
Q-balls~\cite{nts}, with non-zero baryon and lepton numbers~\cite{ak_mssm}.
The interior of these Q-balls has the same field structure~\cite{kst} as
the AD condensate; and, in fact, very large Q-balls can form via the
fragmentation of the scalar condensate in the early Universe~\cite{ks,em}.

Depending on their size and the mode of supersymmetry breaking, the
baryonic balls formed in the process of AD baryogenesis can be stable or
unstable.  If they are stable, they can presently exist as a form of dark
matter~\cite{ks} with distinctive observational signatures~\cite{kkst}.  If
they are unstable~\cite{ccgm}, the LSP produced from their decay can
contribute to dark matter~\cite{em}.  In both cases, the ordinary matter
and the dark matter come from the same origin, the primordial scalar
condensate, and the amounts of dark matter ($\Omega_{_{DM}}$) and the
ordinary matter ($\Omega_{_{M}}$) are related~\cite{em1,ls}.  One may hope
to understand, therefore, why $\Omega{_{_M}}$ and $\Omega_{_{DM}}$ are the
same order of magnitude.  This equality is inevitably fortuitous in
theories that assume the dark matter formed as a result of a freeze-out or
other process unrelated to baryogenesis (see, however, Ref.~\cite{kaplan}).
If, however, both components arise from AD baryogenesis, it is reasonable
to ask whether a relation $\Omega{_{_M}} \sim \Omega_{_{DM}}$ can have a
natural explanation~\cite{em1,ls}.

At the end of inflation, the scalar fields acquire large expectation values
along the flat directions and begin to roll towards their potential
minima.   Under very generic conditions, such a motion of the scalar
condensate can become unstable with respect to small coordinate-dependent
perturbations~\cite{ks}. For an adiabatically slow motion of the initially
homogeneous condensate $\phi(x,t)=\rho(t) \exp(i \Omega(t))$, the
instability sets in when the second derivative of the potential,
$U''(\phi(x,t)) \equiv U''(\rho(t))$, is smaller than $(d\Omega(t)/dt)^2$.
We note in passing that for many potentials, in particular, those that
arise as a result of gauge-mediated supersymmetry breaking,
$U''(\phi(x,t))$ can be negative for some values of $\phi$, in which case
the condition $U''(\phi(x,t))< (d\Omega(t)/dt)^2$ is satisfied
automatically.  If the expansion of the universe occurs at a rate that
allows this instability to grow, the condensate becomes inhomogeneous and
breaks up into Q-balls.  

Q-balls produced from the shattered AD condensate can further evolve by
absorbing baryons from or releasing baryons into the surrounding
plasma~\cite{ls,s_gen}.  

Stable baryonic Q-balls can be the dark matter.  Their
signatures~\cite{kkst} in different detectors are characterized by a
straight track with a large energy deposition, $\sim 100 $~GeV/cm, and no
attenuation throughout the detector volume.  The dark-matter Q-balls would
have typical galactic velocities $\sim 10^{-3}c$.  Assuming an order-one
contribution to energy density of the Universe, one can set limits on their
flux and, hence, their baryon number $Q_{_B}$.  The present limits require
that $Q_{_B}>10^{22}$~\cite{kkst,baikal}.  The predicted range for $Q_{_B}$
consistent with $\Omega{_{_M}} \sim \Omega_{_{DM}}$ is $Q_{_B} = 10^{26 \pm
2}$~\cite{ls}.

The entire cosmologically interesting range of dark-matter
Q-balls~\cite{ks,kkst,ls} can be explored by a detector with a surface
area of several square kilometers.  Since the required sensitivity is
extremely low (thanks to the huge energy release expected from the passage
of a superball), it is conceivable that a relatively inexpensive dedicated
experiment could perform an exhaustive search and ultimately discover or
rule out the stable Q-balls as dark-matter particles.

Baryonic Q-balls can absorb protons and neutrons, a process accompanied by
a release of energy.  This is because a stable baryonic ball has a smaller
mass than a collection of neutrons with the same total baryon number.  As a
result, primordial Q-balls can accumulate in the center of a neutron star,
grow through an absorption of neutrons, and eventually reduce the neutron
star mass below the limit of its stability, which is about $0.2 M_\odot $.
At that point gravity is no longer strong enough to keep the neutrons from
decaying, and the star explodes~\cite{sw}.  The lifetime of a neutron
star is of order $(m/200 {\rm GeV})^5$~Gyr, where $m$ is the scale
associated with supersymmetry breaking.  Perhaps, these mini-supernova
explosions~\cite{minisn}, which release of order $10^{52}$ erg, can be
observable; they may also be related to (some) gamma-ray bursts.

The cosmological limits on dark mater impose some constraints on the form
of the potential along a flat direction responsible for
baryogenesis~\cite{constr}.  Thanks to the sensitivity of the Q-ball
properties  to the high-energy scales ({\it cf.}~\cite{dks}), one may hope
to learn about the higher-dimensional operators due to the Planck-scale
physics, as well as the supersymmetry-breaking terms that lift the flat
directions~\cite{constr}.  

Enqvist and McDonald have suggested that the unstable SUSY Q-balls can also
play a role in the production of dark matter~\cite{em,em1,em2,em3,em4}.
The decay of a SUSY Q-ball into quarks is accompanied by the production of
the lightest supersymmetric particles~\cite{em} that contribute to dark
matter.  This scenario offers an explanation of why $\Omega{_{_M}} \sim
\Omega_{_{DM}}$~\cite{em1} because the dark-matter neutralinos are produced
non-thermally with a number density that is related to the density of
baryons.  This scenario fits naturally in cosmology with a D-term
inflation~\cite{Dterm} and a reheating temperature below $10^{4 \pm
1}$~GeV~\cite{em2}.

Formation and decay of the unstable SUSY Q-balls could cause the isocurvature
fluctuations that may be observable by MAP and PLANK~\cite{em3}.

The scenarios described above were studied in the context of MSSM. Of
course, some additional scalar fields with global charges that often appear
in the non-minimal SUSY models can also form stable Q-balls~\cite{demir}. 
The non-abelian Q-balls~\cite{nonabelian}, as well as the Q-balls
associated with the conservation of a gauge charge~\cite{local}, may also
play an important role in cosmology.

To summarize, the Affleck-Dine baryogenesis in the MSSM can produce both
the matter nucleons and the lumps of the scalar condensate, Q-balls. The
stable Q-balls, as well as the products of the decay of the unstable
Q-balls can contribute to dark matter.  These possibilities are illustrated
by the following diagram:

\vspace{3mm} \centerline{\psfig{figure=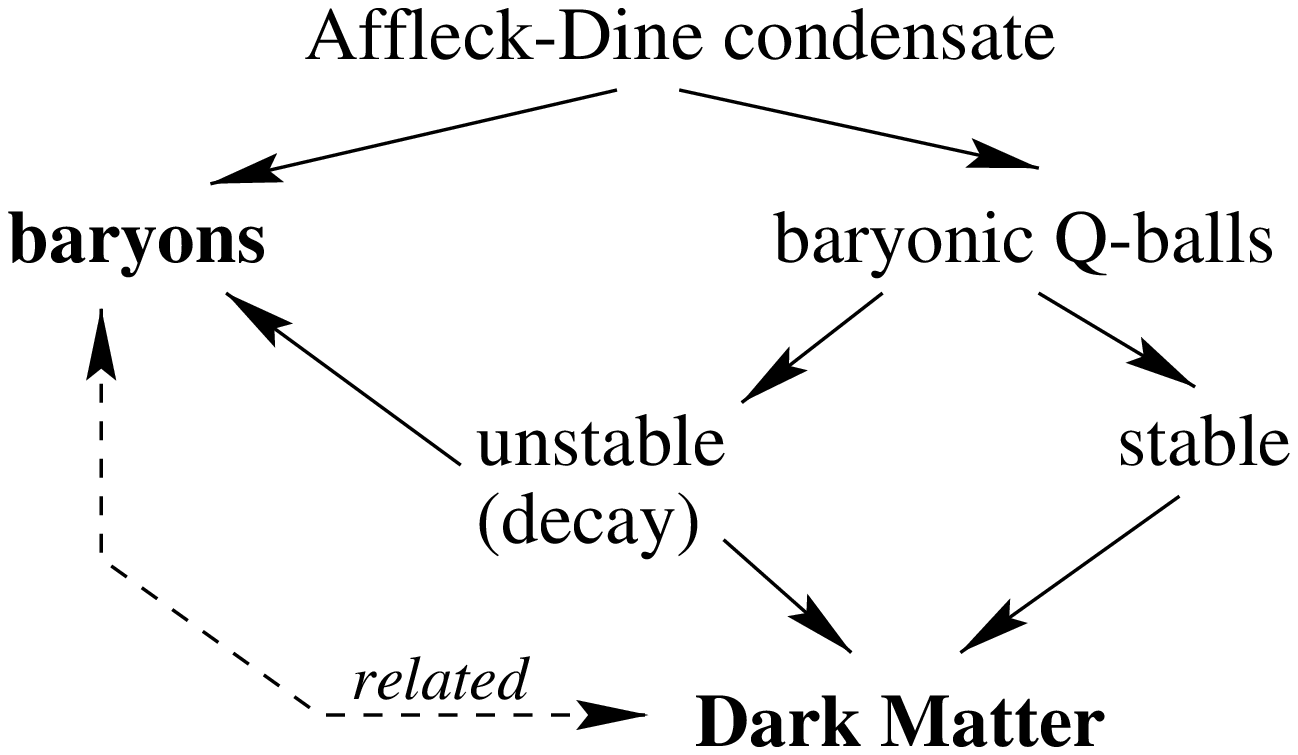,height=50mm,width=110mm}}
Formation of Q-balls can have profound implications for
cosmology\footnote{For a more detailed review, see, {\it e. g.},
Ref.~\cite{ak_rev}.}.  Since the fragmentation of a coherent scalar
condensate~\cite{ks,em} is the only conceivable mechanism that could lead
to the formation of very large Q-balls, their observation would seem to
speak unambiguously in favor of such process having taken place.  Thus,
although non-observation of Q-balls cannot rule out the AD baryogenesis,
their observation would be a definite confirmation thereof.  In addition,
the decay of unstable SUSY Q-balls could produce the LSP dark matter at
late times and out of equilibrium.

\end{document}